\documentclass[pra,twocolumn,superscriptaddress,noshowpacs,longbibliography]{revtex4-1}
\usepackage{amssymb}
\usepackage{amsmath}
\usepackage{amsfonts}
\usepackage{bm}
\usepackage{graphicx}
\usepackage[dvipsnames]{xcolor}
\usepackage{calc} 
\usepackage{epsfig}
\usepackage{epstopdf}
\usepackage{natbib}
\usepackage{hyperref}
\begin{document}

\title{Modeling Collective Behavior of Posting Microblog by Stochastic Differential Equation with Jump}

\author{Jun-Shan Pan}
\affiliation{School of Automation and School of Physics, Huazhong University of Science and Technology, Wuhan, 430074, China}
\affiliation{Key Laboratory of Image Processing and Intelligent Control of Education Ministry, Huazhong University of Science and Technology, Wuhan, 430074, China}

\author{Yuan-Qi Li}
\affiliation{School of Automation and School of Physics, Huazhong University of Science and Technology, Wuhan, 430074, China}
\affiliation{Key Laboratory of Image Processing and Intelligent Control of Education Ministry, Huazhong University of Science and Technology, Wuhan, 430074, China}

\author{Xiang Liu}
\affiliation{School of Automation and School of Physics, Huazhong University of Science and Technology, Wuhan, 430074, China}
\affiliation{Key Laboratory of Image Processing and Intelligent Control of Education Ministry, Huazhong University of Science and Technology, Wuhan, 430074, China}

\author{Han-Ping Hu}
\email{huhp@mail.hust.edu.cn}
\affiliation{School of Automation and School of Physics, Huazhong University of Science and Technology, Wuhan, 430074, China}
\affiliation{Key Laboratory of Image Processing and Intelligent Control of Education Ministry, Huazhong University of Science and Technology, Wuhan, 430074, China}

\author{Yong Hu}
\email{huyong@mail.hust.edu.cn}
\affiliation{School of Automation and School of Physics, Huazhong University of Science and Technology, Wuhan, 430074, China}

\begin{abstract}
  The characterization and understanding of online social network behavior is of importance from both the points of view of fundamental research and realistic utilization. In this manuscript, we propose a stochastic differential equation to describe the online microblogging behavior. Our analysis is based on the microblog data collected from Sina Weibo which is one of the most popular microblogging platforms in China. Especially, we focus on the collective nature of the microblogging behavior reflecting itself in the analyzed data as the characters of the periodic pattern, the stochastic fluctuation around the baseline, and the extraordinary jumps. Compared with existing works, we use in our model time dependent parameters to facilitate the periodic feature of the microblogging behavior and incorporate a compound Poisson process to describe the extraordinary spikes in the Sina Weibo volume. Such distinct merits lead to significant improvement in the prediction performance, thus justifying the validity of our model. This work may offer potential application in the future detection of the anomalous behavior in online social network platforms.
\end{abstract}



\maketitle

\section{Introduction}
\label{Sec Intro}

Online social network has drastically changed the way people behave and interact with others. Through microblogging platforms including Twitter and Sina Weibo, distant users can share information instantaneously when hot topics emerge, resulting in the collective behavior of posting and reposting microblogs. The understanding of such online behavior thus has important potential applications in the administration of network servers, the detection of abnormal behaviors, and the prevention of the spreading of computer viruses. On the other hand, the rapid development of technology has allowed a huge amount of behavior traces recorded and prompted an explosive quantitative research, revealing that online human behavior exhibits complicated patterns of heterogeneity \cite{barabasi2005origin,mathews2017nature,wang2016modeling}, burstness \cite{gandica2016origin,goh2008burstiness,YAN2017775}, memorability \cite{vazquez2007impact,garcia2017memory,gleeson2016effects}, and periodicity \cite{zhou2012relative,yan2012human,vazquez2006impact}. Correspondingly, a variety of models have been proposed based on the analyses of the behaviors on the individual level, which attributed the observed phenomena to the mechanisms of task driven \cite{barabasi2005origin,vazquez2006impact}, interest driven \cite{han2008modeling,zhao2013emergence}, circadian rhythm \cite{malmgren2008poissonian,hidalgo2006conditions} or others \cite{zhao2016dynamic,jo2015correlated,tavares2013scaling}.

Meanwhile, the online social network behavior has intrinsically collective nature the understanding of which is of significant importance and has already become the focus of recent research. In particular, Mathiesen et. al \cite{mathiesen2013excitable} have found out that the power spectral density and the broad distribution observed in the tweet rate are related to the strong correlation on a global scale in the collective human dynamics and proposed a consequent model of stochastic point process. Also, Mollgaard et. al \cite{mollgaard2015emergent} have suggested that the strongly correlation of the tweet behavior can be extracted from the tweet rate including a particular brand name and can be modeled by a stochastic different equation (SDE), with a linear drift term describing the inactive behavior without any external input and a diffusion term characterizing the burst behavior trigged by external events. However, these proposed stochastic models still need further discussion in the sense that whether the linear drift term used in these models is capable of grasping the ubiquitous periodic pattern observed in the behavior of Twitter and Sina Weibo, and whether the existing few but extremely stiff spikes in behavior data can be effectively characterized by the diffusion term.

The aim of this manuscript is to develop an alternative SDE model with improved functionality in the characterization and prediction of the online social network behavior. Our analysis and modeling is based on the microblog data collected from Sina Weibo, which is one of the most popular online social media platforms in China. Similar to Twitter, users of Sina Weibo can share information and viewpoints by posting microblogs limited in $140$ words, and their followers can comment on the original microblogs or repost them to their own homepages. The number of monthly active users of Sina Weibo has exceeded $340$ million by year 2016, making it an ideal platform of investigating the online social network behavior. Explicitly speaking, we study the collective behavior by analyzing the time series of Sina Weibo volume data aggregated by all the collected users. Such time series is characterized by its properties of regular periodicity, stochastic volatility and extreme jumps. We then propose a SDE model to describe the analyzed microblogging behavior. Rather than the fixed constants in existing stochastic models which prevent the emergence of periodical patterns, here we introduce the time dependent parameters to guarantee the mean reversion process towards a periodic baseline. In addition, we incorporate a compound Poisson process into the proposed SDE to depict the irregular spikes of the analyzed time series which are trigged by some extraordinary events. Our numerical benchmark indicates that these two distinct merit can significantly improve the prediction performance of the SDE model, thus justifying its validity and opening up a possible route toward the thorough cognition of online social behavior.

This manuscript is organized as follows. We first describe in Sec. \ref{Sec Data} the method of data collection and then analyze empirically the properties of corresponding microblogging behavior in Sec. \ref{Sec Empirical}. The proposed SDE model is descried in detail in Sec. \ref{Sec SDE} with its performance discussed in Sec. \ref{Sec Performance}. Finally, we conclude our work and offer outlook in Sec. \ref{Sec Conclusion}.


\section{Data description}
\label{Sec Data}

In this manuscript, we focus on the collective properties of the microblogging behavior on Sina Weibo. The data for our analysis is collected through the Sina Weibo API by using a web crawler software. Here we exploit the snowball sampling algorithm \cite{snowball} which starts from a random user and crawl along the following and followed relation in an iterate manner. Finally we collect $38,554,643$ microblogs from $977,546$ users during Jan. 1. 2015 to Dec. 31. 2015 as our dataset. As the comments appear only as the attachments of the individual microblogs, in this collection we only take the original microblogs and the repost ones into account. We then obtain the Sina Weibo volume data (SWVD) time series by counting the number of microblogs per time unit, which is analyzed hereafter. Such SWVD time series is a typical representation of the dynamic characteristic of collective behavior since its fluctuation is produced by all users of Weibo. In addition, we choose $1$ hour as our unit time scale, which is much longer than the $1$ second or $1$ minute time scale used in the analysis of web surfing behavior. This is due to the empirical observation that the microblogging behavior is much more sparse than web surfing.

\begin{figure*}[tbh!]
\begin{center}
\includegraphics[width=0.95\textwidth]{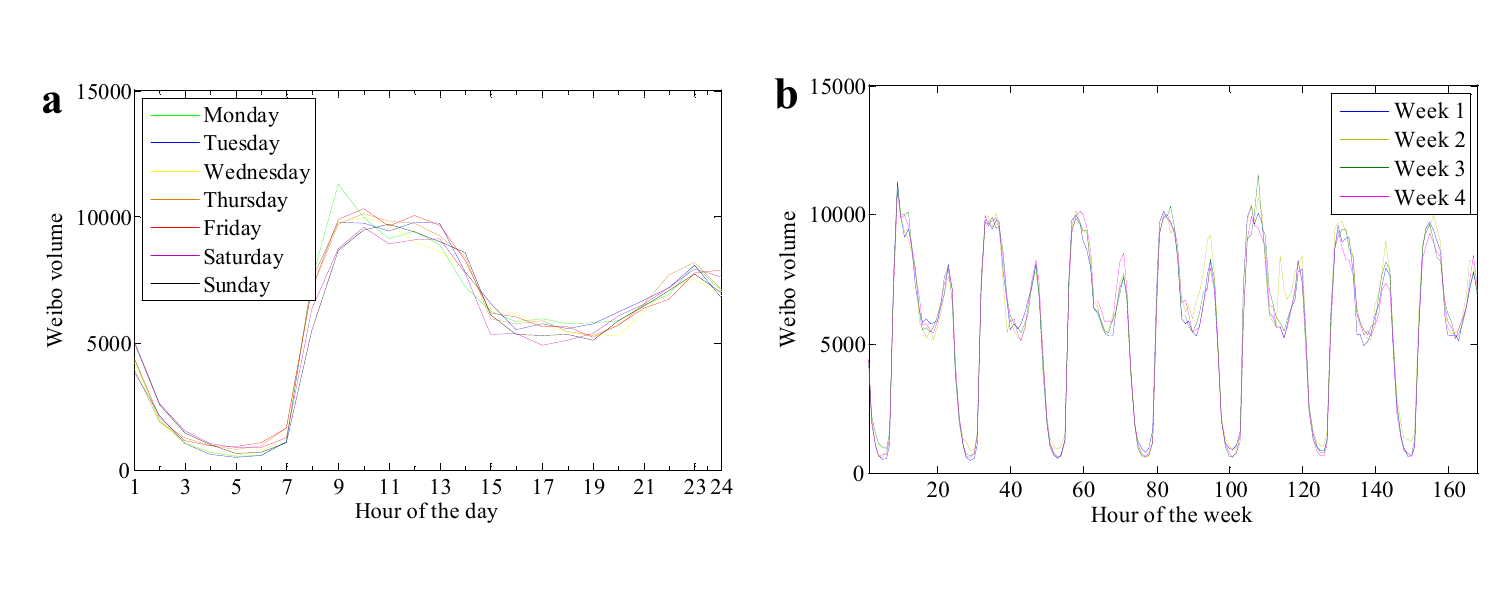}
\end{center}
\caption{\label{Fig Period} Periodic pattern of the SWVD series. The daily fluctuation of the SWVD series during a randomly chosen week is shown in (a), while the weekly fluctuation during a randomly chosen month is depicted in (b).}
\end{figure*}

\section{Empirical analysis}
\label{Sec Empirical}

Human behavior is regarded to be the result of the interaction between external environment and internal factors of human beings. This statement is represented by Lewin's equation of behavior \cite{lewin2013principles}
\begin{align}
\label{Eqn Lewin}
  B=f\left( P, E \right),
\end{align}
where $B$ is the variable characterizing a certain kind of human behavior, $P$ represents the internal factors of human beings, and $E$ describes the external environment. Following this criteria, we analyze the Weibo posting behavior from two aspects, i. e.  the internal factor and external environment.

\begin{figure*}[tbh!]
\begin{center}
\includegraphics[width=0.95\textwidth]{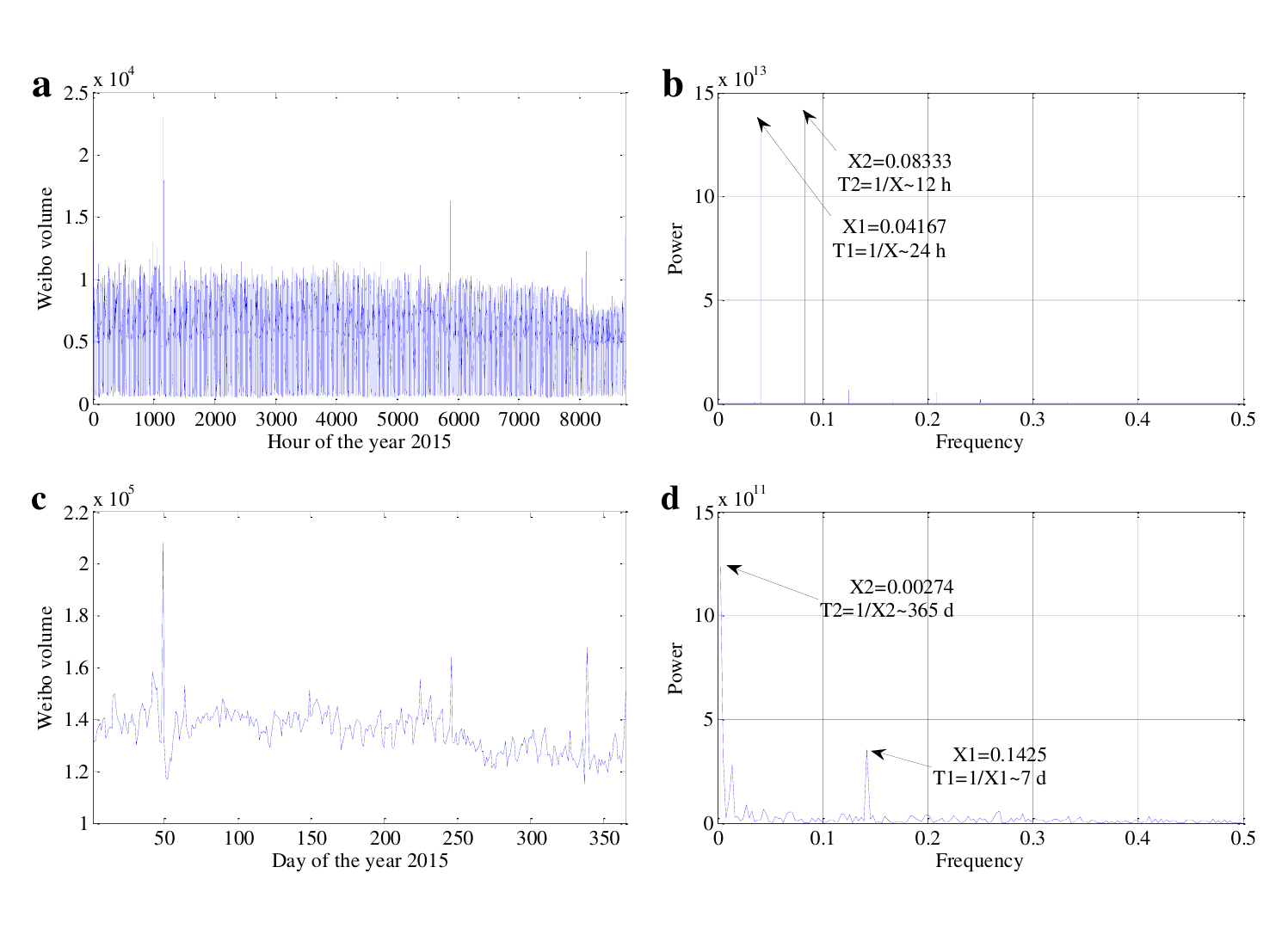}
\end{center}
\caption{\label{Fig FFT} Fluctuation of the SWVD during the whole year of 2015 in time resolution of (a) $1$ hour and (c) $1$ day. The corresponding power spectra obtained from FFT is shown in (b) and (d), respectively, with peaks indicating the periods of the SWVD series.}
\end{figure*}

\textit{The intrinsic factors.---} To demonstrate the influence of the intrinsic factors on the microblogging behavior, we plot in the first step the daily fluctuation of the SWVD time series during a randomly chosen week in Fig. \ref{Fig Period}(a), with each colorline corresponding to a single day in the week. Similar to other online behaviors, the periodic property of the analyzed microblogging behavior can be clearly observed: The SWVD tends consistently to rise up during the daytime and drop down during the nighttime. Such $1$--day periodic pattern is clearly the reflection of the circadian rhythm of human, as most people are likely to wake up and perform certain jobs during 8--23 and then go to sleep during 0--7. In addition, the valleys of the Weibo volume data appear almost at the same positions around 12 and 18 everyday, which correspond to the lunch and dinner times, respectively. The fast Fourier transformation (FFT) technique \cite{Hassani} is then applied to extract the periodic property of the SWVD series: The SWVD fluctuation during the whole 2015 year in time resolution of $1$ hour is shown in Fig. \ref{Fig FFT}(a), and the corresponding power spectrum given by FFT is plotted in Fig. \ref{Fig FFT}(b), with two peaks indicating the $24$--hour and $12$--hour periods being clearly discerned.

We go further to analyze the weekly of the SWVD. In Fig. \ref{Fig Period}(b), the weekly fluctuation of the SWVD during a randomly chosen month is plotted with each colorline corresponding to one of the four weeks in the month, where the $1$--week period pattern can be clearly observed. However, such $1$-week period does not appear in the power spectrum shown in Fig. \ref{Fig FFT}(b). The reason is that such weekly periodic pattern is less distinct than the $24$--hour period in the time resolution of $1$ hour. Meanwhile, the $1$--week cycle is recovered when we scale the time resolution down to $1$ day: The SWVD fluctuation in time scale of $1$ day is depicted in Fig. \ref{Fig FFT}(c), while the corresponding power spectra obtained from FFT is shown in Fig. \ref{Fig FFT}(d), with the peak representing the $1$--week cycle emerged. The existence of such $1$--week period can also be interpreted in an intuitive way: most people tends to schedule their appointment in the scale of a week and then perform different activities during each day of the week, e. g. they work in the weekdays and relax in the weekends. From this point of view, it is more reasonable to compare behaviors in different Mondays than to compare behaviors in Monday and Sunday. Based on this observation, we take $1$ week rather than $1$ day as our time window in the later modeling of the microblogging behavior.

\textit{The external environment.---} Affected by the circadian rhythm, people tend to develop behavior habbits, i. e. to repeat behavior in the same period of a day or a week. On the other hand, the random fluctuation of the analyzed behaviors are also observed, which is induced by various intrinsic and external factors. Let us take the lunch behavior as an example. Suppose 12 is the regular lunch time, people can rarely able to have lunch at 12 precisely every day. Instead, the lunch behavior will be performed with a random fluctuation around 12. This intuitive observation reflects that human behavior should be regarded as a stochastic fluctuation process around its baseline, which is also called regression toward the average in the field of social psychology \cite{myers1987social}. Such random fluctuation has already manifested itself in Fig. \ref{Fig Period}: The daily and the weekly records of the SWVD coincide not completely but within a very small fluctuating range.

\begin{figure*}[tbh!]
\begin{center}
\includegraphics[width=0.95\textwidth]{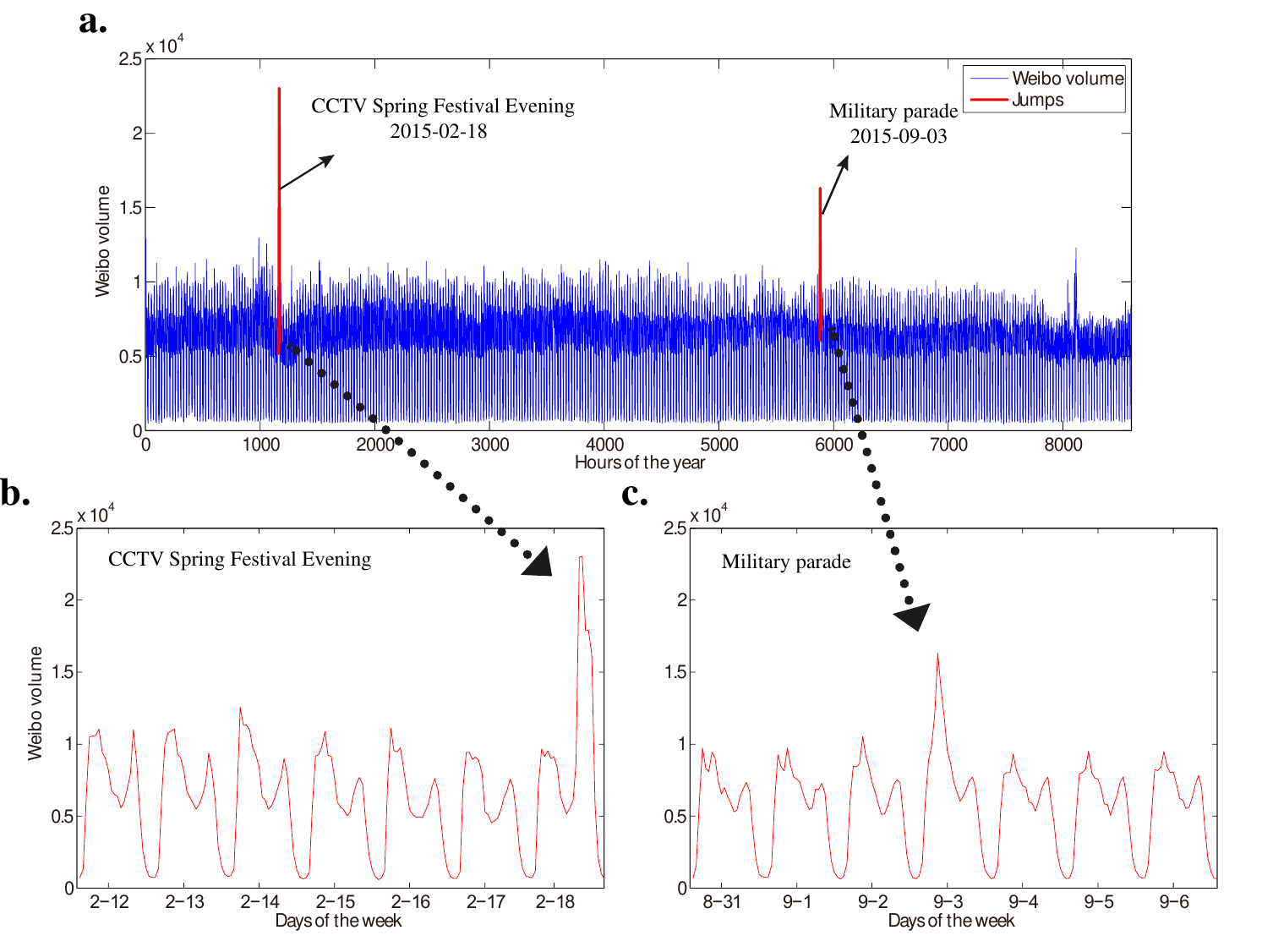}
\end{center}
\caption{\label{Fig Burst} Jumps in the SWVD. Two jumps marked by the red peaks are caused by the 2015 CCTV Spring Festival Evening on Feb. 18. 2015 and the anti-Japanese War victory parade of China on Sep. 3. 2015, respectively. The corresponding bursts of the SWVD around the jumps are detailed in  (b) and (c).}
\end{figure*}

Moreover, there are few extraordinary spikes appeared, e. g. at 9 of Monday in Fig. \ref{Fig Period}(a) and at the 110th hour of week 2 in Fig. \ref{Fig Period}(b). On the Sina Weibo platform, hot topics including scheduled big events, disasters, and emergency public events can attract attentions of many users in a very short time. In these circumstances, individuals will share the topic-related information by reposting the original news or posting their own comments, and with the wide spreading of these microblogs, increasing users will be involved into the topics, resulting in irregular bursts of the SWVD denoted by jumps. The occurrence of  jumps can significantly deviate the behavior of the SWVD from its own baseline. For instance, there is an extraordinary peak located at Feb. 18. 2015 of Figs. \ref{Fig Burst}(a) and (b), which is related to the CCTV Spring Festival Evening 2017, the biggest entertainment evening of China celebrating the Chinese new year (this relation is verified by tracing back these Weibo events through a heat curve of hot events application provided by Sina Weibo). In addition, another notable peak is located at Sep. 03. 2015. This jump is induced by the anti--Japanese War victory parade of China (Fig. \ref{Fig Burst}(c)).

Following the above analysis, we come to the summary that the SWVD series have reflected a few important collective patterns of the microblogging behavior, namely the regular period, the stochastic fluctuation around the baseline, and the jumps. The modeling of these mentioned properties is thus the main task of the remainder of this manuscript.

\section{SDE model of the SWVD}
\label{Sec SDE}

In this section, we aim at developing a SDE model of describing the microblogging behavior on the Sina Weibo platform. In particular, our attention is focused on the properties of period, stochastic volatility, and jumps of the SWVD. The volatility of the SWVD is caused by the collective behavior of many users, and such collective motion will result in anomalous jump patterns when certain hot spots happen. On the other hand, the similar situation has also been encountered in the area of economics, where the fluctuations of stock prices and interest rates can be regarded as the consequence of the combined action of various investors. A variety of stochastic models including Black--Scholes (BS) \cite{black1973pricing} and Cox--Ingersoll--Ross (CIR) \cite{cox1985theory} have been proposed to model these complicated financial behaviors successfully. In particular, the CIR model can well describe the properties of mean reversion and positive definite of the interest rate behavior. Moreover, being ubiquitous in the field of economics,  jump has been modeled by compound Poisson process, and various phenomenon including  ``volatility smile'' and ``leptokurtosis and fat-tail''  can only be effectively simulated with such jump term taken into consideration. Inspired by the advances in the field of economics, we  propose in the following a modified CIR model with jump to characterize the collective microblogging behavior reflected by the SWVD.

\begin{figure*}[tbh!]
\begin{center}
\includegraphics[width=0.95\textwidth]{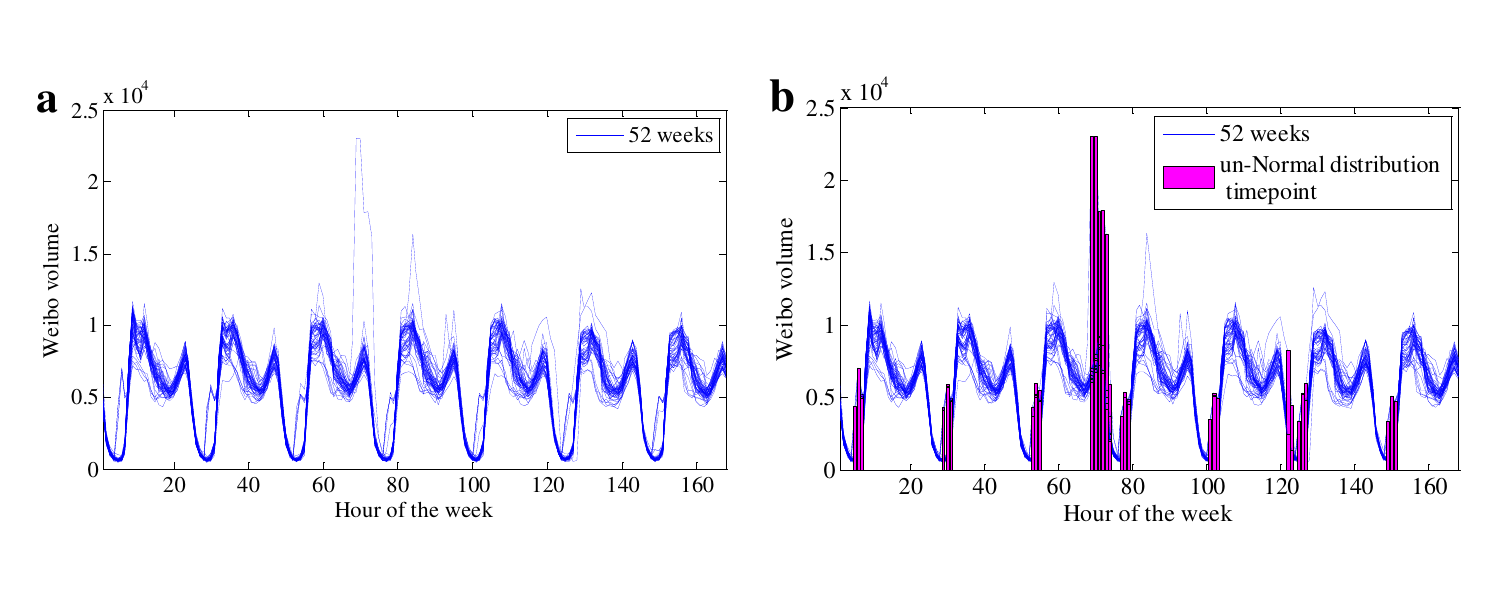}
\end{center}
\caption{\label{Fig Detection} SWVD fluctuation during the 52 weeks of year 2015 in time resolution of $1$ hour. $31$ of the $168$ vertical time series do not follow Normal distribution, with the corresponding time points marked by purple bars in (b).}
\end{figure*}

\textit{Jump Detection.---} As discussed previously, the jump phenomenon influences the behavior of the SWVD significantly despite its rare occurrence. Therefore, the first step of our modeling is to detect and analyze the jumps in the SWVD in a quantitative way. Existing methods of jump detection in economic research such as wavelet transform \cite{wang1995jump} and the nonparametric method \cite{lee2007jumps} are suitable only for stationary financial data. However, the SWVD is quite different in its periodic property, which causes false positive detection of jump around the local maximum of the SWVD in a cycle (e. g. the lunch time shown in Fig. \ref{Fig Period}(a)). Considering the periodic pattern and the mean reversion property of Weibo behavior, we propose an alternative vertical comparison detecting method to determine the jumps: We choose the time window as $1$ week and form $168$ vertical time series (corresponding to the $168$ hours of a week) by extracting the SWVD at the same time point of each week (Fig. \ref{Fig Detection}(a)); The baseline of each vertical time series is defined as the calculated average of this vertical time series, and an anomalous deviation from the calculated baseline is regarded as a jump. Here we should be careful that the trend presented in the SWVD should be eliminated before performing the above tasks: If the original SWVD has a e. g. decreasing trend, data in the first few weeks of the year may be false positively identified as jumps. Therefore, we detrend the SWVD before detecting jumps by first fitting out the linear trend of the original SWVD with the least square method and then subtracting this linear trend from the original data (notice that the de-trended data should be remained positive-definite as the SWVD can naturally not be negative).

\begin{figure*}[tbh!]
\begin{center}
\includegraphics[width=0.95\textwidth]{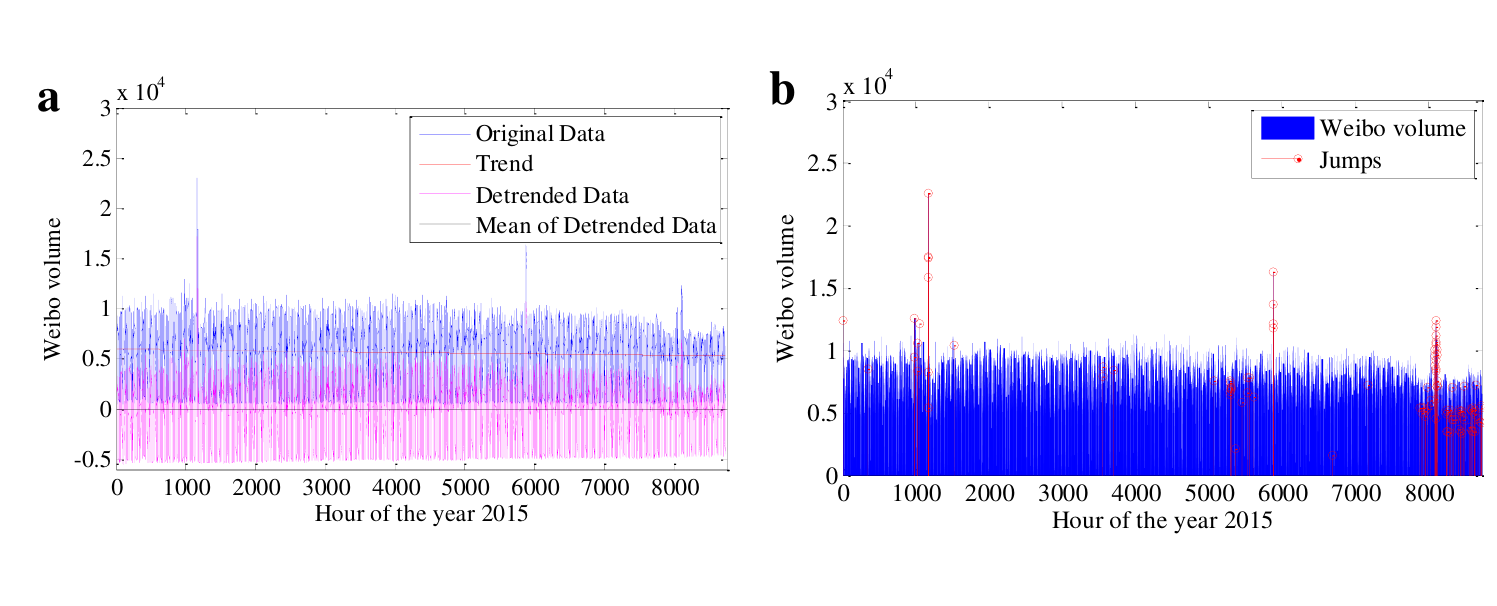}
\end{center}
\caption{\label{Fig Detrend} (a) SWVD of year 2015 in time resolution of $1$ hour. The original and the detrended data are labeled by the blue and the purple lines, respective. For the clarity of demonstration the detrended data are moved downward globally. The trend of the original data and the mean value of the detrended data are marked by the red dashed and the black dashed lines, respectively. The jumps in the analyzed SWVD is determined by using the method described in the main text and denoted by the circles in (b).}
\end{figure*}

Following the strategy described above, we first plot in Fig. \ref{Fig Detrend}(a) both the original SWVD in the year 2015 and the corresponding detrended data. We then investigate the distribution of each vertical time series of the de-trended data through Kolmogorov-Smirnov (KS) test \cite{fellerbook1,fellerbook2}. Our result shows that $137$ of the $168$ vertical time series follow Normal distribution, with the time points corresponding to the remaining 31 series marked by purple bars in Fig. \ref{Fig Detection}(b). The extreme studentized deviation (ESD) identification method \cite{rosner1983percentage} can therefore be further employed for the jump detection: Suppose we have a SWVD of $n$ weeks. Denoting $X_i(t)$ the SWVD value at the time point $t$ in the $i$th week, we identify $X_i(t)$ as a jump if the following condition is fulfilled:
\begin{align}
\label{Eqn Jump}
  \left| X_i(t)-  \overline{X(t)}  \right| \geq \kappa\sigma(t),
\end{align}
where
\begin{align}
\label{Eqn Parameter}
  \overline{X(t)}=\sum_{i=1}^{n} \frac{X_i(t)}{n},\qquad\sigma(t)=\sqrt{\frac{\sum_{i=1}^{n} \left[  X_i(t)-  \overline{X(t)}  \right]^2  }{n-1}},
\end{align}
are the baseline at the time point $t$ and the corresponding standard deviation for $t \in \left[ 1, 168\right]$, respectively. In addition,  the parameter $\kappa=3$ can be assigned because the probability of observing values outside the $3\sigma$ range of Normal distribution is less than $0.3\%$ \cite{fellerbook1}. The finally obtained jump detection result is sketched in Figure. \ref{Fig Detrend}(b).

\textit{Modeling the collective behavior of the SWVD.---} The CIR model which inspires our present research has the SDE form
\begin{align}
\label{Eqn CIR}
  \mathrm{d}X(t)=k\left[ a-X(t) \right] + c\sqrt{X(t)}\mathrm{d}W(t),
\end{align}
where the first term at the righthand side is the drift term ensuring the mean reversion towards the longrun value $a$ with $k$ being a strictly positive parameter controlling the speed of adjustment, and the second term is the diffusion term avoiding the possibility of negative $X(t)$ value for positive $k$ and $a$ with $c$ being the diffusion constant and $W(t)$ being a Wiener process. Nevertheless, the CIR model in Eq. (\ref{Eqn CIR}) can not be exploited directly to model the collective behavior of the SWVD because the constant parameters of the model fail to grasp the periodic pattern of the SWVD. What is more, Eq. (\ref{Eqn CIR}) dose not consider the jumps of the SWVD. Our solution is to propose a improved CIR model, taking the form
\begin{align}
\label{Eqn ModifiedCIR}
  \mathrm{d}X(t)=&k\left[ a(t)-X(t) \right]  \notag \\
  &+ c(t)e\sqrt{X(t)}\mathrm{d}W(t)+J(t)\mathrm{d}N(t),
\end{align}
where a new third term describing the jumps in the SWVD has been incorporated with $N(t)$ being a Poisson process with parameter $\lambda (t)$ and $J(t)$ being the amplitude of the jumps. In addition, the parameters $a$ and $c$ in Eq. (\ref{Eqn CIR}) are updated as time dependent variables and a new constant $e$ is introduced. The setting of the parameters in Eq. (\ref{Eqn ModifiedCIR}) are base on their physical meaning explicitly explained below:

\begin{figure}[tbh!]
\begin{center}
\includegraphics[width=0.48\textwidth]{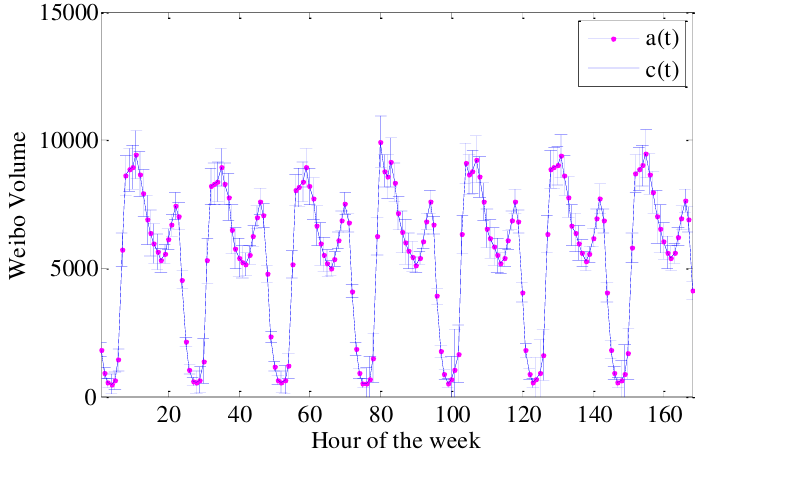}
\end{center}
\caption{\label{Fig BaseLine} Calculated baselines and standard deviations of the vertical time series.}
\end{figure}

\begin{itemize}
  \item The drift term $k\left[ a(t)-X(t) \right]$ sketches the regression towards the average of the SWVD series with $a(t)$ being the target mean value. Based on the periodic pattern of the Weibo volume data described before, we no longer set the target mean value as a constant (see the original CIR model in Eq. (\ref{Eqn CIR})) but treat it as a periodic function with its period being $1$ week and its value being the baseline of the vertical time series at that time point, i. e. $a(t)=\sum^{n}_{i=1}X_i(t)/n$ (here it should be noticed that $X_i(t)$ here are the values of the detrended SWVD with the jumps removed, see the later determination of the jump parameters). The calculated $a(t)$ based on the SWVD of year 2015 is shown in Fig. \ref{Fig BaseLine}. In addition, the constant parameter $k$ (along with the parameter $e$ described in the following) is estimated by using Euler method \cite{EulerMethod} after all the time-dependent parameters are determined.
  \item The diffusion term $c(t)e\sqrt{X(t)}\mathrm{d}W(t)$ induces stochastic fluctuation of $X(t)$ deviating from its baseline. The parameter $e$ controls the amplitude of the stochastic fluctuation and $c(t)$ denotes the volatility of $X(t)$ at time point $t$. Therefore, we set $c(t)$ as the standard deviation of vertical time series at time point $t$, i.e. $c(t)=\sqrt{{\sum_{i=1}^{n} \left[  X_i(t)-  a(t)  \right]^2  }/{(n-1)}}$. Along with the previously mentioned $a(t)$, the calculated $c(t)$ based on the SWVD of 2015 is also shown in Fig. \ref{Fig BaseLine}
  \item The jump term $J(t)\mathrm{d}N(t)$ represents the influence of the jump events on the behavior of the SWVD. Inspired by advances in economics, we model the jump of the Weibo volume data as a compound Poisson process with $\lambda(t)$ denoting the possibility of jump appearance at time point $t$ and $J(t)$ describing the amplitude of the jump. In particular, $\lambda(t)$ can be determined by the number $m_{\mathrm{count}}(t)$ of jumps detected at time point $t$ during the whole $n$ weeks, i. e. $\lambda(t)=m_{\mathrm{count}}(t)/n$. Based on the previous results of the Normal distribution property of the SWVD, we further set that $J(t)$ follows Normal distribution with mean value $\mu(t)$ and standard deviation $\sigma(t)$. The parameters $\mu(t)$ and  $\sigma(t)$ can then be directly extracted from the detected jumps of the SWVD.
\end{itemize}

\section{Performance of the proposed model}
\label{Sec Performance}

\begin{figure*}[tbh!]
\begin{center}
\includegraphics[width=0.95\textwidth]{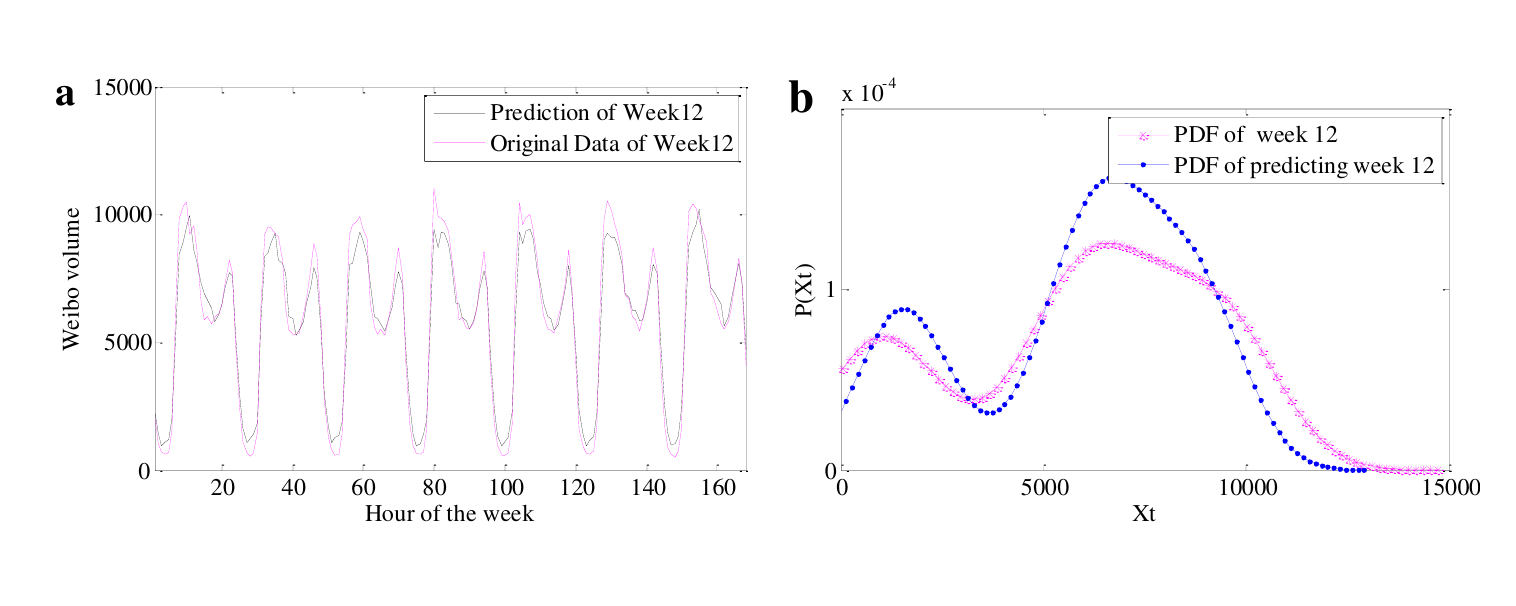}
\end{center}
\caption{\label{Fig Prediction} Performance of the proposed model in predicting the SWVD in the $12$th week. The predicted and the real SWVD trajectories are plotted in (a), while the corresponding PDF are depicted in (b).}
\end{figure*}

\begin{figure*}[tbh!]
\begin{center}
\includegraphics[width=0.95\textwidth]{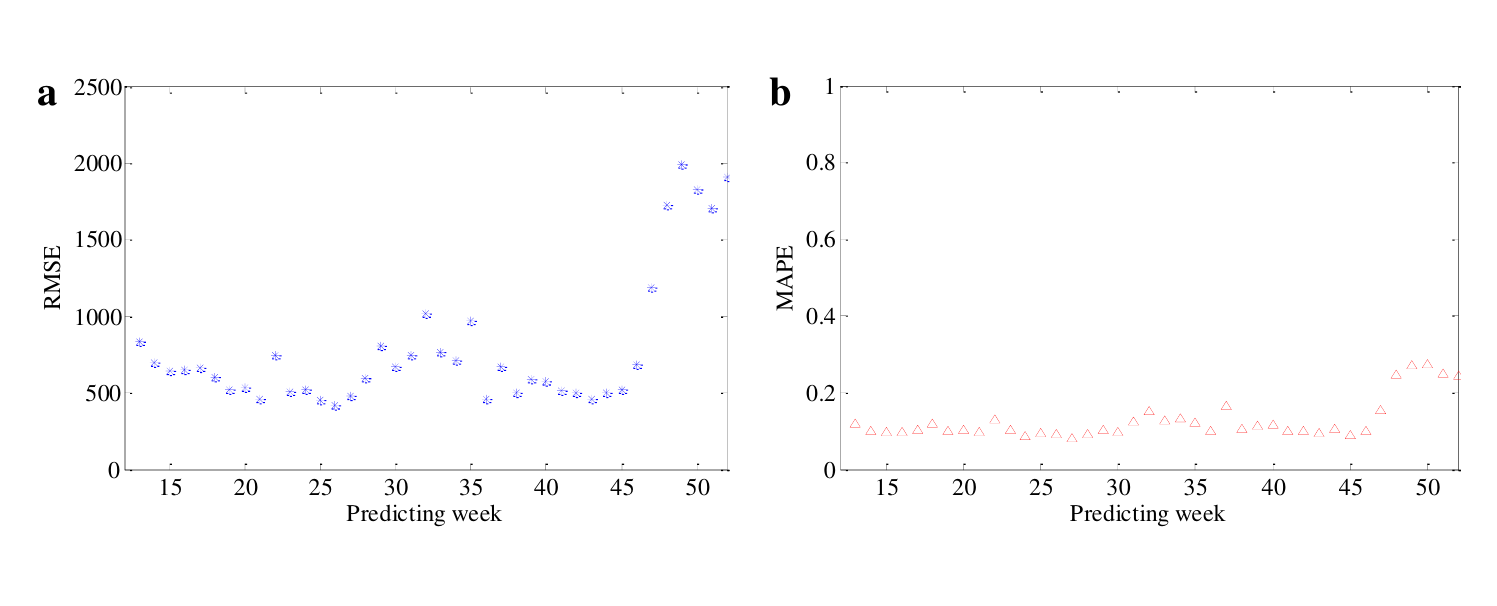}
\end{center}
\caption{\label{Fig PredictError} Performance of the proposed model in predicting the SWVD in the $12$th--$52$th weeks. The prediction performance are characterized by RMSE and MAPE shown in (a) and (b), respectively.}
\end{figure*}

\textit{Performance of predictivity.---} In this section we discuss the performance and validity of our proposed model from a variety points of view. The whole $52$--week SWVD of year 2015 is applied to test the prediction performance of our model. Explicitly speaking, in the first step we use the SWVD of the first $11$ weeks as the training data to estimate the parameters of our model and then we exploit the estimated parameters to simulate the SWVD trajectory of the $12$th week. Here the jumps in the training data are removed based on the method illustrated in the previous section, and the length of the training data is set due to the research conducted by psychologists Philippa Lally which found out that the average time required for a behavior to become a habit is $66$ days \cite{LallyHabits}. We then obtain the predicted trajectory of the SWVD in the $12$th week shown in Fig. \ref{Fig Prediction}(a) by numerically generating $1000$ stochastic trajectories from the proposed SDE and calculating their average. The prediction performance of the proposed model is quantified by two widely used indicators of prediction error, namely, the root mean square error (RMSE) and the  mean absolute percentage error (MAPE), which measure the absolute error and the relative error, respectively. These indicators takes the form
\begin{align}
  \mathrm{MAPE}&=\frac{1}{L}\sum_{t} \left| \frac{X_{\mathrm{real}}(t)-X_{\mathrm{pre}}(t)}{X_{\mathrm{real}}} \right|, \notag\\
  \mathrm{RMSE}&=\sqrt{\frac{1}{L}{\sum_{t}\left[ X_{\mathrm{real}}(t)-X_{\mathrm{pre}}(t)  \right]^{2}}},
\end{align}
where $X_{\mathrm{real}}(t)$ and $X_{\mathrm{pre}}(t)$ are the measured real data and the data predicted by the proposed model, respectively, $L$ is the length of the predicted data, and the summations run over the whole time for prediction (e. g. the $12$th week). The MAPE and RMSE of the predicted data are $19.6\%$ and $625.8$, respectively. To further visualize the performance of the model, the probability density function (PDF) of the predicted and the real Weibo volume data over the $12$th week is calculated and plotted in Fig. \ref{Fig Prediction}(b). By employing KS test, we verify that the predicted and the corresponding actual trajectories follow the same probability distribution at the confidence level of $5\%$. From these demonstrated prediction performance, we thus come to the conclusion that the validity of our model is justifies. We further generalized the above process to the whole analyzed SWVD of the $52$ weeks, that is, we use the data in the first $N$ weeks as training data to estimate the parameters and generate the prediction trajectory of the $(N+1)$th week for $11 \leq N \leq 51$. The corresponding MAPE and RMSE between the real and the predicted data from week $12$ to week $52$ are depicted in Figs. \ref{Fig PredictError}(a) and (b), respectively, suggesting that our model can characterize various properties of the microblogging behavior effectively.

\begin{figure*}[tbh!]
\begin{center}
\includegraphics[width=0.95\textwidth]{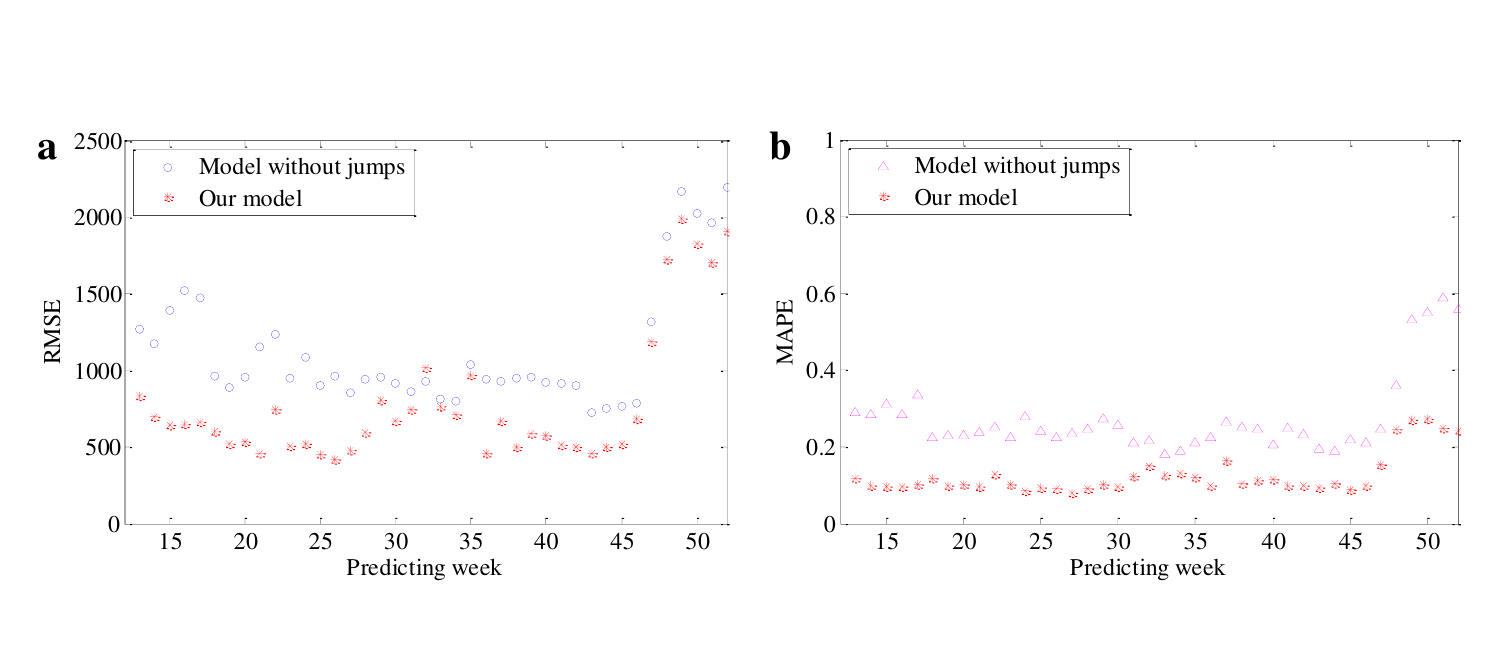}
\end{center}
\caption{\label{Fig WithoutJump} Comparison of the prediction performance of the models with and without the jump term. The corresponding RMSE and MAPE in the predicted $12$th--$52$th weeks are shown in (a) and (b), respectively.}
\end{figure*}

\textit{Comparing with the model without jump.---} Jump events are very important because their appearance can deviate the behavior of the SWVD severely from its baseline. From this point of view, it is the distinct merit of this manuscript that we take the jumps into consideration when developing our model in Eq. (\ref{Eqn ModifiedCIR}). To further illustrate the advantage of the jump term in our model, we compare the results obtained from  Eq. (\ref{Eqn ModifiedCIR}) with the results from the following model
\begin{align}
\label{Eqn WithoutJump}
  \mathrm{d}X(t)=k^{\prime}\left[ a^{\prime}(t)-X(t) \right] + c^{\prime}(t)e^{\prime}\sqrt{X(t)}\mathrm{d}W(t),
\end{align}
i. e. Eq. (\ref{Eqn ModifiedCIR}) with the jump term removed. Here the parameters  $a^{\prime}(t)$, $c^{\prime}(t)$, $k^{\prime}$ and $e^{\prime}$  are estimated by using the similar method mentioned in the previous section. The prediction of the SWVD during the $12$th to the $52$th weeks are then re-performed, with the corresponding RMSE and MAPE results shown in Figs. \ref{Fig WithoutJump}(a) and (b), respectively. Compared with the results given by Eq. (\ref{Eqn ModifiedCIR}), the rising up of the RMSE and MAPE calculated from Eq. (\ref{Eqn WithoutJump}) can be identified, suggesting again the importance of considering the jumps in characterizing the behavior of the SWVD. The difference in the predictive performance between these two models can be traced back to the jumps existing in the training weeks (i. e. the first $11$ weeks of the analyzed data), which have significant influence on the estimation of time dependent parameters $a(t)$ and $c(t)$. In particular, by using the without-jump model in Eq. (\ref{Eqn WithoutJump}), we would overestimate the value of $c(t)$ and in turn generate a more severe fluctuation in the subsequent numerical simulation.

\begin{figure*}[tbh!]
\begin{center}
\includegraphics[width=0.96\textwidth]{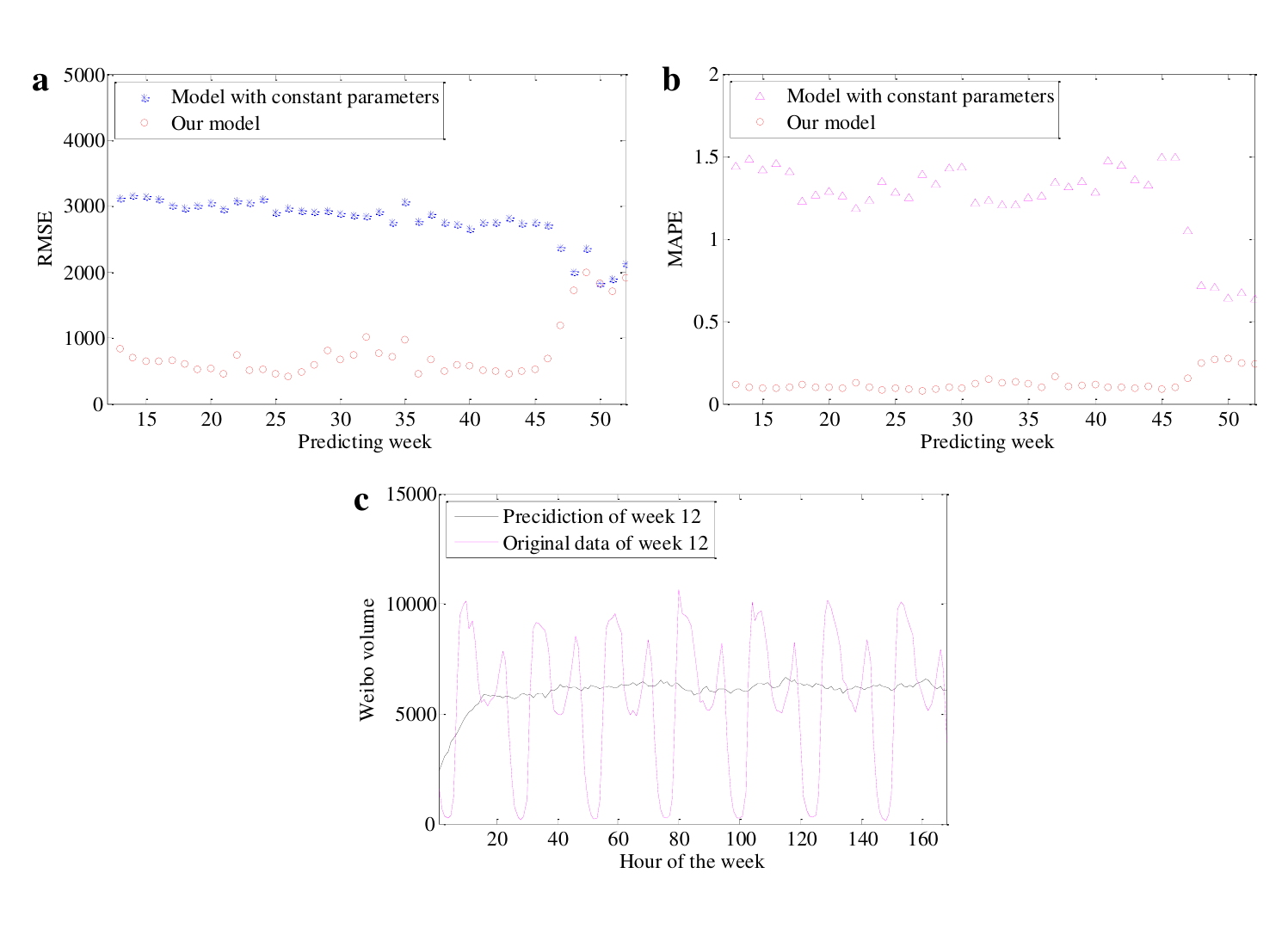}
\end{center}
\caption{\label{Fig ConstantParameter} Prediction performance of the models with constant and time dependent parameters. The corresponding RMSE and MAPE in the predicted $12$th--$52$th weeks and the predicted trajectories of the $12$th week are shown in (a), (b), and (c), respectively.}
\end{figure*}

\textit{Comparing with the model with constant parameter.---} The substitution of constants by time dependent parameters is the distinction between our model and the original CIR model. This improvement is to facilitate the periodical behavior of the SWVD, and its advantage can be further illustrated by comparing it with the following constant-parameter model
\begin{align}
\label{Eqn ConstantParameter}
  \mathrm{d}X(t)=k\left[ a^{\prime\prime}-X(t) \right] + c^{\prime\prime}\sqrt{X(t)}\mathrm{d}W(t)+J^{\prime\prime}\mathrm{d}N(t).
\end{align}
The consequent jump detection, parameter estimation, and the prediction of SWVD behavior can be re-performed similar to the process discussed previously. As shown in Figs. \ref{Fig ConstantParameter}(a) and (b), the resulting RMSE and MAPE are much higher than those given by Eq. (\ref{Eqn ModifiedCIR}), indicating the advantage of introducing the time dependent parameters. To trace back the difference of the prediction performance, we plot in Figure. \ref{Fig ConstantParameter}(c) the baseline of the trajectory predicted by Eq. (\ref{Eqn ConstantParameter}), which suggests that such constant-parameter model fails to generate the periodic patterns of the SWVD behavior. This incapability is also intuitive: the physical interpretation of $a(t)$ in Eq. (\ref{Eqn ModifiedCIR}) is the average value of the SWVD which exhibits periodic property. However, it has been simplified to a constant  in Eq. (\ref{Eqn ConstantParameter}). It is this unphysical simplification that leads to the deviating results shown in Fig. \ref{Fig ConstantParameter}.

\section{Conclusion and Outlook}
\label{Sec Conclusion}
In conclusion, we have developed in this manuscript a SDE model to describe the observed collective patterns of the online microblogging behavior including regular period, stochastic volatility, and jumps. The distinct merit of this scheme is that we introduce time dependent parameters and jump term, which significantly improve the prediction performance of the proposed model. Therefore, our work may offer an alternative route towards the understanding of the online social network behavior, and its potential application in determining the baseline of the normal behavior as well as detecting consequently the anomalous behavior in online social network platforms should be our future research direction.

\begin{acknowledgments}
This work was supported in part by the National Science Foundation of China (Grants No.~11374117 and No.~11774114) and the Knowledge Innovation Program for Basic Research Project of Shenzhen (Grant No. JCYJ20150616144425386).
\end{acknowledgments}

%



\end{document}